\documentclass[10pt,twocolumn,showpacs,amsmath,amssymb,floatfix,superscriptaddress]{revtex4-2}

\usepackage{graphicx}
\usepackage{dcolumn}
\usepackage{bm}
\usepackage{xcolor}
\usepackage{txfonts}
\usepackage{microtype}
\usepackage{epsfig}
\usepackage{hyperref}
\usepackage[normalem]{ulem}

\begin{document}	
\title{Generation of Ultrabrilliant Positron Beam via  Superponderomotive Injection in Laser Wakefield Acceleration}

\author{Ting Sun}
\affiliation{Ministry of Education Key Laboratory for Nonequilibrium Synthesis and Modulation of Condensed Matter, State key laboratory of electrical insulation and power equipment, Shaanxi Province Key Laboratory of Quantum Information and Quantum Optoelectronic Devices, School of Physics, Xi'an Jiaotong University, Xi'an 710049, China}
\affiliation{School of Physics, Nankai University, Tianjin 300071, China}	
\author{Zhen-Ke Dou}
\affiliation{Ministry of Education Key Laboratory for Nonequilibrium Synthesis and Modulation of Condensed Matter, State key laboratory of electrical insulation and power equipment, Shaanxi Province Key Laboratory of Quantum Information and Quantum Optoelectronic Devices, School of Physics, Xi'an Jiaotong University, Xi'an 710049, China}	
\author{Ya-Qing Huang}
\affiliation{Ministry of Education Key Laboratory for Nonequilibrium Synthesis and Modulation of Condensed Matter, State key laboratory of electrical insulation and power equipment, Shaanxi Province Key Laboratory of Quantum Information and Quantum Optoelectronic Devices, School of Physics, Xi'an Jiaotong University, Xi'an 710049, China}	
\author{Feng Wan}
\affiliation{Ministry of Education Key Laboratory for Nonequilibrium Synthesis and Modulation of Condensed Matter, State key laboratory of electrical insulation and power equipment, Shaanxi Province Key Laboratory of Quantum Information and Quantum Optoelectronic Devices, School of Physics, Xi'an Jiaotong University, Xi'an 710049, China}	
\author{Qian Zhao}\email{zhaoq2019@xjtu.edu.cn}
\affiliation{Ministry of Education Key Laboratory for Nonequilibrium Synthesis and Modulation of Condensed Matter, State key laboratory of electrical insulation and power equipment, Shaanxi Province Key Laboratory of Quantum Information and Quantum Optoelectronic Devices, School of Physics, Xi'an Jiaotong University, Xi'an 710049, China}	
\author{Jian-Xing Li}\email{jianxing@xjtu.edu.cn}
\affiliation{Ministry of Education Key Laboratory for Nonequilibrium Synthesis and Modulation of Condensed Matter, State key laboratory of electrical insulation and power equipment, Shaanxi Province Key Laboratory of Quantum Information and Quantum Optoelectronic Devices, School of Physics, Xi'an Jiaotong University, Xi'an 710049, China}	
\affiliation{Department of Nuclear Physics, China Institute of Atomic Energy, P.O. Box 275(7), Beijing 102413, China}

\date{\today}
	
\begin{abstract}
Plasma-based acceleration of positrons attracts extensive interest owing to the ultrahigh accelerating gradient and ultrashort duration, while generating wakefield positron beam by the inherent injection is still a great challenge. Here, we put forward a superponderomotive injection method of positrons in the blowout regime of laser wakefield acceleration.
{\color{black}{The dephasing-rate integral equation reveals a twofold mechanism: the longitudinal laser field delays phase-locking, guiding positrons into the paraxial focusing region, while the transverse laser Lorentz force suppresses the dephasing rate below unity, trapping them into the laser-modulated wakefield. Particle-in-cell (PIC) simulations demonstrate this via a donut-wake--pair-jet collision, generating low-emittance ($\sim$0.05~mm~mrad) multicycle positron beams. Start-to-end simulations for post-acceleration in the second-stage donut wakefield confirm high-throughput injection-to-acceleration coupling, yielding quasi-monoenergetic beam with six-dimensional brightness $\sim 10^{15}~\rm{A/m^2}/0.1\%$. This plasma-based injection-acceleration scheme opens a novel compact route to ultrabrilliant positron sources for ultrafast material diagnostics, laboratory astrophysics, and next-generation electron--positron colliders.}}
\end{abstract}

\maketitle
Generating relativistic positron beams with ultrabrilliance stands as a pivotal endeavor of particle physics, high-energy astrophysics and material science with profound implications \cite{cao2024Positron,si2024Research}. 
Beyond the ongoing development of an electron-positron Higgs factory \cite{an2019Precision,adolphsen2022European,narain2023Future,farmer2024Preliminary,geng2024Compact}, relativistic positron beams are well-suited for investigating laboratory astrophysics by modeling the pair-driven extreme-radiation astrophysical environment \cite{uzdensky2014Plasma}, such as pulsar magnetospheres \cite{arrowsmith2024Laboratory}, black hole jets\cite{ruffini2010Electron}, and gamma-ray bursts \cite{kumar2015physics}. Especially, ultrafast positron beams with tunable keV-MeV energy have transformative potential for probing quantum phase transitions and material defects by annihilation lifetime spectroscopy \cite{Keeble2010identification, audet2021ultrashort} and microscopy \cite{david2001lifetime}. Traditionally, the sophisticated trap-based  technologies have been designed to generate positron beam, based on the positron sources by radioisotopes in nuclear reactors and LINAC-based electron-positron ($e^\pm$) pairs \cite{danielson2015Plasma,fajans2020Plasma}. 
Despite the cumbersome facilities, relatively low trapping efficiency, and limited acceleration gradient inherent to radio-frequency accelerators, trap-based technologies remain challenged in delivering positrons as an ultrabrilliant, femtosecond-scale beam \cite{hessami2023Compact}.

Plasma wakefield acceleration (PWFA) enables micron-scale accelerator to  produce ultrabrilliant particle beams, owing to its ultrahigh accelerating gradients ($\gtrsim100$ GV/m) and intrinsic femtosecond duration (less than a quarter of plasma wavelength) \cite{wang2016HighBrightness,li2022Ultrabright,wan2023Femtosecond,fuchs2024Plasma}. Meanwhile, high-power laser enables highly efficient $e^\pm$ pair production via bremsstrahlung-induced Bethe-Heitler (BH) process in a high-$Z$ converter target \cite{sarri2015Generation,chen2015Scaling,chen2023Perspectives}, opening the pathway to create the compact relativistic positron source \cite{doche2017Acceleration,alejo2019Laserdriven,streeter2024Narrow}.
\textcolor[rgb]{0.00,0.0,0.00}{With the development of PWFA schemes capable of positron acceleration---including the vanishing transverse field in hollow channel plasma and the positron-focusing wakefield driven by hollow laser---alongside laser-driven BH sources reaching pair densities of $\sim10^{18}$ cm$^{-3}$ \cite{chen2023Perspectives}, numerous investigations have been devoted to the external injection of a pre-tailored positron beam into a hollow plasma wakefield \cite{kimura2011Hollow,yi2013Scheme,schroeder2013Control,gessner2016Demonstration,silva2021Stable,zhou2021High}, or a donut wake \cite{jain2015Positron, vieira2014Nonlinear,yu2014Control,firouzjaei2017Trapping}.} The external injection method requires  stringent spatiotemporal properties (micron-scale emittance and femtosecond-scale duration) of a trailing positron beam for the sake of injection into a plasma wakefield \cite{corde2015Multigiga,hessami2023Compact}. Such a requirement significantly challenges the current extracting and collimating technologies (e.g., via a magnetic spectrometer) for generating a positron beam from BH $e^\pm$ pairs with broad divergence ($10^\circ-40^\circ$) and relativistic energy (corresponding to effective temperature about 1-20 MeV) \cite{streeter2024Narrow}.
Except for the self-loaded wakefield acceleration which is driven by the evolution of a pre-accelerated positron beam or pair jet with GeV energy in a plasma \cite{corde2015Multigiga,silva2023Positron},
no inherent injection method has hitherto been demonstrated by which energetic positrons from a pair jet can be trapped and accelerated inside the blowout wake.

\begin{figure*}[t!]
\setlength{\abovecaptionskip}{0.2cm}  	
\centering\includegraphics[width=0.95\linewidth]{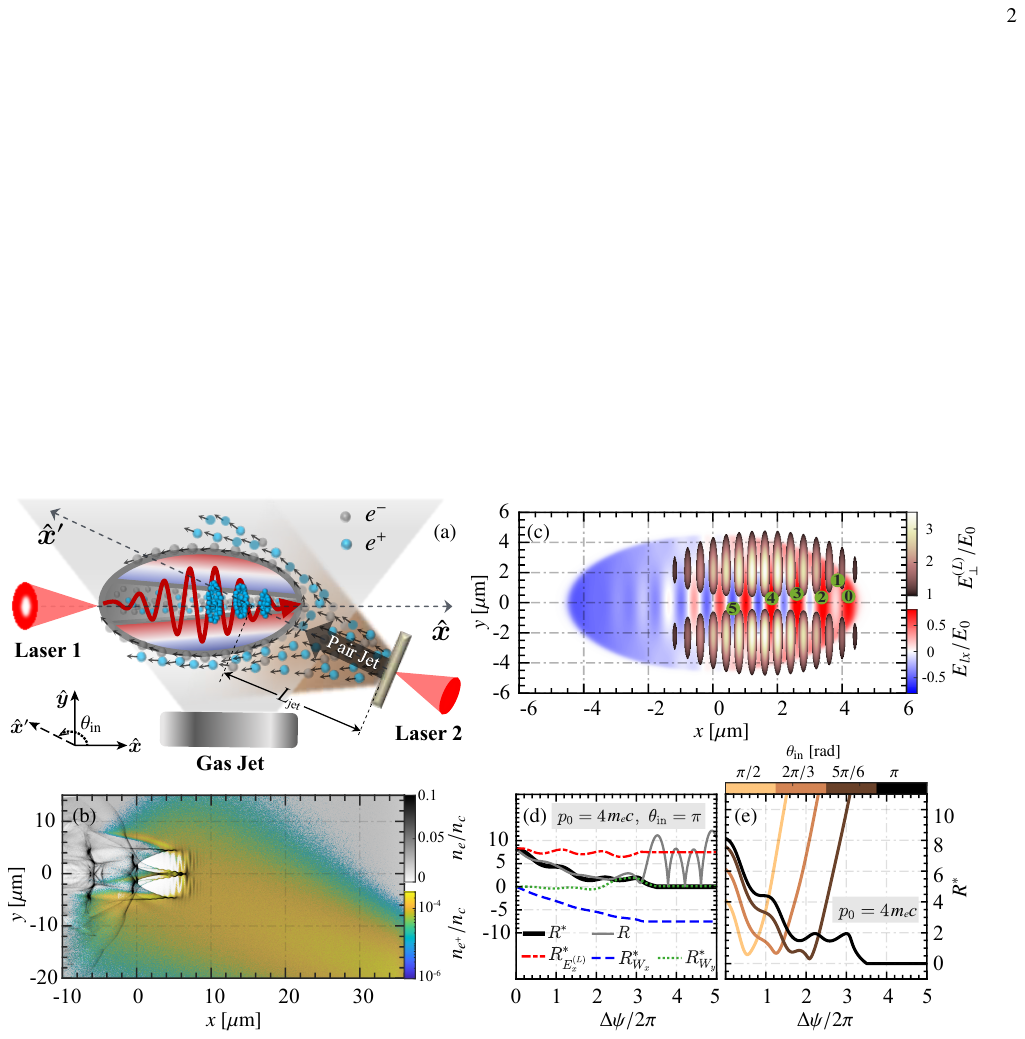}
\caption{{\color{black}(a) Sketch of wake-jet collision configuration: a hollow Laser 1 beam drives a donut wake with transverse wakefield $W_y$ (blue-red shade) and longitudinal laser field $E_x^{(L)}$ (red-solid line); Laser 2 (picosecond duration) impinges on a high-$Z$ converter to generate a relativistic pair jet via the BH process. (b)  3D PIC simulation of wake-jet collision: snapshot of plasma electron density $n_{e}$ and positron density $n_{e^+}$ of a pair jet with incident angle $\theta_{\mathrm{in}}=5\pi/6$. (c) Test-particle model with laser-modulated longitudinal wakefield $E_{lx}$ overlaid by a transverse laser field $E_\perp^{(L)}$. Green circles trace the instant positions of head-on incident positions, labeled with initial momenta $p_0=(0,1,2,3,4,5)~m_ec$. (d) Variation of $R$ and $R^*$ with phase delay $\Delta\psi$ for a positron with $p_0=4~m_ec$ and $\theta_{\mathrm{in}}=\pi$, calculated from Eqs. (\ref{dR_exact}) and (\ref{eq_Rpsi}), respectively, together with components of $R^*$. (e) $R^*$ for positrons with fixed $p_0=4~m_ec$ and varying $\theta_{\rm{in}}=(\pi/2,2\pi/3,5\pi/6,\pi)$.}}
\label{fig1}
\end{figure*}
{\color{black}Directly capturing positrons from a laser-driven pair jet into a donut wake would dramatically simplify PWFA-based positron acceleration. Drawing on the superponderomotive electron mechanism---where an electrostatic field together with laser field reduce the dephasing rate $R=\gamma-p_x/m_ec$ below unity, enabling energy gain from laser and longitudinal momenta exceeding the ponderomotive limit $m_eca_0^2/2$ \cite{arefiev2012Parametric,robinson2013Generating,sorokovikova2016Generation,singh2022Vacuum}---a donut-wake--jet collision should trigger an analogous process for positrons: the annular wakefield inherently funnels positrons toward the axis while expelling electrons radially, and once positrons enter the paraxial region, the combined laser and wakefield suppress $R$, trapping them into the accelerating phase. Here, $\gamma$ and $p_x$ are the Lorentz factor and longitudinal momentum, respectively; $m_e$ and $c$ the electron mass  and speed of light in vacuum, respectively; $a_0$ the normalized laser amplitude.}

\begin{figure}[!t]	
	\setlength{\abovecaptionskip}{0.2cm}  	
	\centering\includegraphics[width=0.95\linewidth]{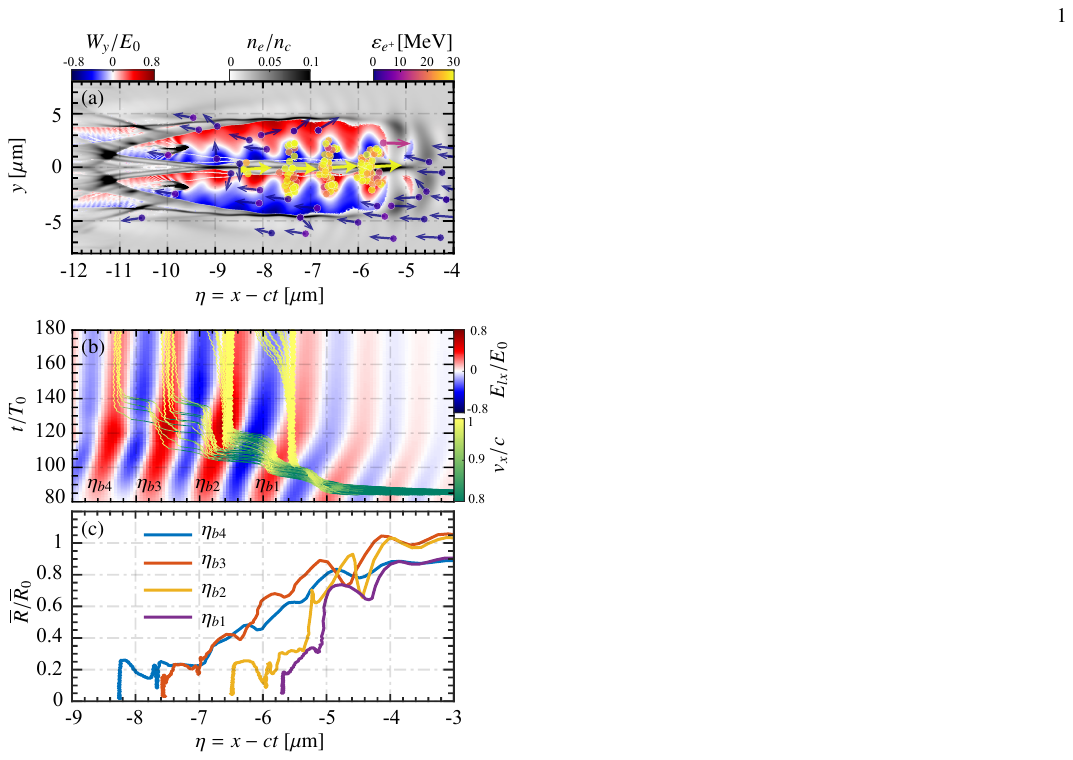}
	\caption{{\color{black}(a) Snapshot of transverse wakefield $W_y$ (blue-red) overlaid by plasma electron density $n_e$ (gray) in the $\eta$--$y$ plane at $t=50~T_0$. Representative positron macro-particles are annotated with instantaneous velocity vectors and energies $\varepsilon_{e^+}$; yellow arrows denote the mean momentum direction of phase-locked positrons. (b) Evolution of the longitudinal field $E_{lx}(y=0)$ in the $\eta$--$t$ plane, overlaid by positron trajectories color-coded by instantaneous longitudinal velocity $v_x$. Positrons are injected into distinct accelerating phases, labeled $(\eta_{b1},\eta_{b2},\eta_{b3},\eta_{b4})$. (c) Evolution of the averaged dephasing rate $\overline{R}$ normalized by averaged initial one $\overline{R_0}$ for the positron bunches trapped in different laser phases identified in (b).}}
	\label{fig2}
\end{figure}

\begin{figure}[!t]
	\setlength{\abovecaptionskip}{0.2cm}
	\centering\includegraphics[width=0.95\linewidth]{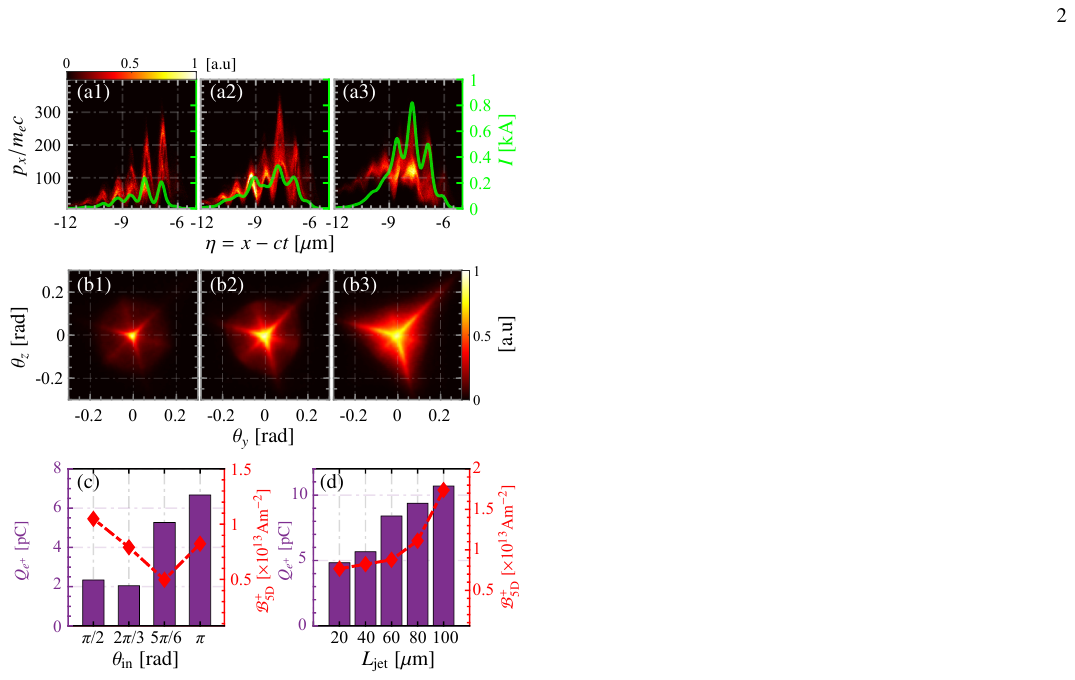}\\
	\caption{{\color{black}(a1)-(a3) Normalized intensity distribution of injected positrons in $p_x$--$\eta$ plane, overlaid by green-lined current intensity $I$,  at fixed $L_{\rm{jet}}=40~\mu\mathrm{m}$ and $\theta_{\rm{in}}=\pi$, for plasma densities $n_{0}=0.025~n_c,~0.035~n_c,~0.05~n_c$, respectively. (b1)-(b3) Same as (a1)-(a3) but for normalized intensity distribution in $\theta_y$--$\theta_z$ plane. \textcolor[rgb]{0.00,0.00,0.00}{3D PIC simulation results at $n_{0}=0.05\,n_c$ for beam charge $Q_{e^+}$ and five-dimensional brightness $\mathcal{B}_{5\mathrm{D}}^+$ versus (c) $\theta_{\rm{in}}$ at fixed $L_{\rm{jet}}=40~\mu$m, and (d) $L_{\rm{jet}}$ at fixed $\theta_{\rm{in}}=5\pi/6$.}} }
	\label{fig3}
\end{figure}

\begin{figure}[!t]
	\setlength{\abovecaptionskip}{0.2cm}
	\centering\includegraphics[width=0.95\linewidth]{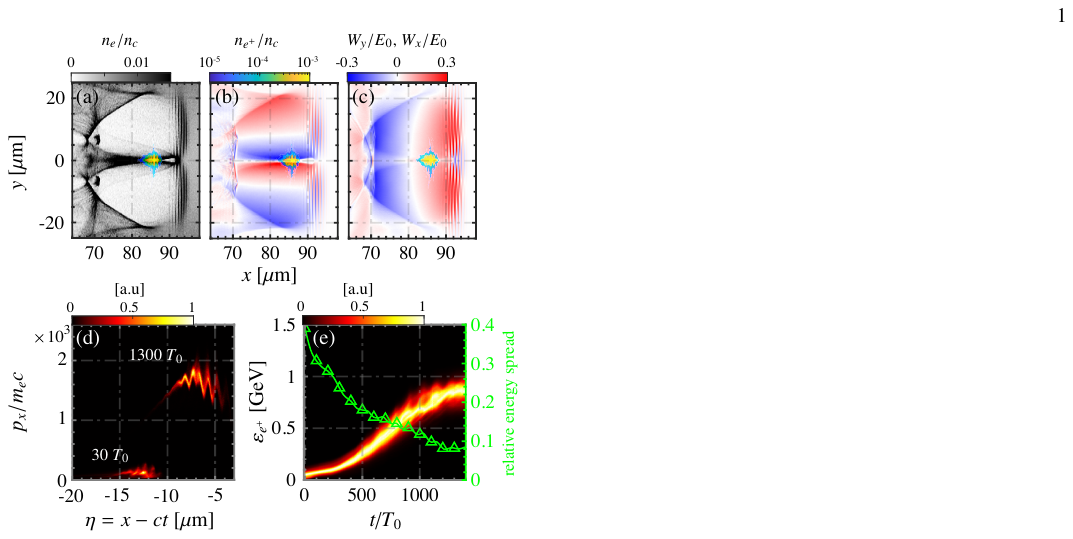}\\
	\caption{{\color{black}Post-acceleration of the trailing positron beam from Fig.~3(a3) in an LG-laser-driven wakefield. (a) Superposition of plasma density $n_e$ and trailing beam density $n_{e^+}$ at $t=30\,T_0$. (b) Superposition of transverse wakefield $W_y$ and  $n_{e^+}$. (c) Superposition of longitudinal wakefield $W_x$ and $n_{e^+}$. (d) Phase-space distribution of the trailing beam at time $t=30\,T_0$ and $t=1300\,T_0$. (e) Temporal evolution of the trailing beam spectrum; overlayed by the relative energy spread (green solid-scattered line).} }
	\label{fig4}
\end{figure}

{\color{black}In this Letter, we propose a superponderomotive injection method that enables inherent trapping of BH-jet positrons in the blowout regime of laser wakefield acceleration (LWFA). The dephasing-rate integral equation, Eq.~(\ref{eq_Rpsi}), reveals a mechanism whereby the longitudinal laser field $E_x^{(L)}$ delays phase-locking, guiding positrons into the paraxial focusing region, while the Lorentz force of the transverse laser fields suppresses $R$ below unity, trapping them into the laser-modulated wakefield $E_{lx}=E_x^{(L)}+W_x$, where $W_x$ is the longitudinal wakefield. This scheme is realized via the collision between a donut wake and a relativistic BH pair jet produced at a high-$Z$ converter [Figs.~\ref{fig1}(a)-(b)]. 
A test-particle model predicts the sensitivity of injection to the incident angle [Figs.~\ref{fig1}(c)–(e)]. Particle tracking in three-dimensional (3D) PIC simulation confirms that, as divergent positrons are captured into the paraxial focusing field and longitudinal accelerating phases, their dephasing rate converges toward zero---the hallmark of superponderomotive injection in laser fields [Fig.~\ref{fig2}]. 
Scans over plasma density, incident angle, and collision distance demonstrate robust injection, generating low-emittance multicycle positron beams [Fig.~\ref{fig3}]. Start-to-end simulations confirm high-throughput injection-to-acceleration coupling: post-acceleration of the trailing beam in a second-stage donut wakefield yields quasi-monoenergetic positron beam with a six-dimensional brightness of $\sim10^{15}~\mathrm{A\,/m^{2}}/0.1\%$ [Fig.~\ref{fig4}].} 

\paragraph*{Proposal of superponderomotive injection---}In the proposed scheme, a donut wake driven by a tightly focused hollow pulse (Laser~1) propagates along $\hat{\bm x}$ and collides at an angle $\theta_{\rm in}$ with a relativistic pair jet propagating along $\hat{\bm x}'$ [Fig.~\ref{fig1}(a)]. Laser~2 (picosecond, kilojoule-class) impinges on a high-$Z$ converter, generating the pair jet via the BH process along the target normal \cite{chen2015Scaling}; \textcolor[rgb]{0.00,0.0,0.00}{the collision length $L_{\rm jet}$ is the distance from the converter rear surface to the intersection of the $\hat{\bm x}$ and $\hat{\bm x}'$ axes.}
{\color{black}The configuration of Fig.~\ref{fig1}(a), implemented via 3D PIC (parameters in Fig.~\ref{fig2}), funnels entering positrons toward the axis by annular $W_\perp$---which simultaneously expels plasma electrons radially---while those reaching the paraxial region can be captured and accelerated by $E_{lx}$ [Fig.~\ref{fig1}(b)]. Laser fields overlap the quasi-blowout wake bubble, producing a laser-modulated donut wakefield that governs positron dynamics [Fig.~\ref{fig1}(c)] \cite{lu2006nonlinear,sun2024Generation,supplement}. }
\textcolor[rgb]{0.00,0.0,0.00}{This collision configuration serves as a positron injection stage and proceeds on a 100s~femtosecond, far shorter than the 1--10~picosecond BH pair-jet duration \cite{chen2023Perspectives}, substantially relaxing the synchronization requirement between the wakefield and the pair jet.}
{\color{black}Hereafter $a_l=\omega/\omega_p$, time and electric field are normalized to $T_0=2\pi/\omega$ and $E_0=m_ec\omega/e$, respectively, with $\omega$ the laser frequency.}

{\color{black}To distill the superponderomotive mechanism, we employ a test-particle model in the $x$--$y$ plane. A radially polarized laser (RPL) pulse in the paraxial approximation can be simplified by a constant phase velocity $v_{\rm{ph}}$ \cite{salamin2006Accurate}: $E_r^{(L)} = E_{\perp}(r) \cos(\omega(t - x/v_{\rm{ph}}))$, $E_x^{(L)} = E_{\parallel}(r)G(r)\sin(\omega(t - x/v_{\rm{ph}}))$, and $B_\phi^{(L)} \simeq E_rc/v_{\rm{ph}}$; the radial envelope functions $E_{\perp}(r)$, $E_{\parallel}(r)$ and $G(r)$ are given in the Supplemental Material \cite{supplement}. The donut wake is modeled by $W_{\perp} = f(r)k_p(r-r_m)/2a_l$ and $W_x = f(r)k_p(x - r_b - v_bt)/2a_l$ with $f(r)=[\tanh(r_b-r)-1]/2$  \cite{vieira2014Nonlinear,zhang2015Synergistic,hidding2006Generation,kostyukov2009Electron,li2014Dependence}, where $r_b$ and $v_b \simeq c$ are bubble radius and velocity, respectively, $k_p=\omega_p/c$, and $r_m\simeq w_0/\sqrt{2}$ the betatron center.}

{\color{black}From the differential equations of motion $\dot p_x=E_x^{(L)}+W_x+v_yB_z^{(L)}$ and $\dot\gamma=v_x\dot p_x/m_ec^2+v_y\dot p_y/m_ec^2$, the dephasing rate $R=\gamma-p_x/m_ec$ evolves as
\begin{equation}\label{dR_exact}
\dot R=(v_x-c)\,(E_x^{(L)}+W_x)/c+v_y(E_y^{(L)}+W_y-B_z^{(L)})/c.
\end{equation}
With $E_y^{(L)}\simeq B_z^{(L)}$ (valid for $v_{\rm ph}\simeq c$) and $1-v_x/c=R/\gamma$, Eq.~(\ref{dR_exact}) reduces to $\dot R^{*}=-(R^*/\gamma)(E_x^{(L)}+W_x)+v_y W_y/c$, where $R^*$ denotes $R$ evaluated with $v_{\rm ph}=c$. Using $\dot\psi=R/\gamma$, this yields
\begin{equation}\label{dRdpsi}
\frac{dR^*}{d\psi}=-(E_x^{(L)}+W_x)+\frac{\gamma}{R^*}\frac{v_y W_y}{c},
\end{equation}
which integrates to
\begin{equation}\label{eq_Rpsi}
R^*(\psi)=R_0+\!\int_{\psi_0}^{\psi}\!\Big[E_\parallel G\sin\psi'-W_x+\frac{\gamma}{R^*}\frac{v_y W_y}{c}\Big]d\psi'.
\end{equation}
The three integrand terms define the partial contributions $R^*_{E_x^{(L)}}$, $R^*_{W_x}$, and $R^*_{W_y}$. Comparing $R$ and $R^*$ isolates the effect of the superluminal phase velocity ($v_{\rm ph}>c$) accumulated over the phase delay $\Delta\psi=\psi-\psi_0$.}

The test-particle simulation employs an RPL pulse with $\tau_L=5~T_0$, spot size $w_0=3~\mu$m, and peak power $P=10$~TW, and a plasma density $n_e=10^{-3}~n_c$ ($n_c$ is critical density corresponding to laser wavelength $\lambda=0.8~\mu$m). 
\textcolor[rgb]{0.00,0.0,0.00}{Under these parameters, the blowout wake is excited by $\tau_L<2\sqrt{a_0}/\omega_p$ with $a_0=E_\perp(r_m)/E_0\simeq3.5$, and $E_{lx}$ develops a modulated structure within the donut wake [Fig.~\ref{fig1}(c)].}
{\color{black}For head-on collision ($\theta_{\rm in}=\pi$), positrons with different initial momenta $p_0$ are trapped in distinct accelerating phases [numbered circles in Fig.~\ref{fig1}(c)]. In the approximation $v_{\rm ph}=c$, $R^*(\psi)$ reveals that anti-dephasing ($R^*<1$) is dominated by the longitudinal contributions $R^*_{W_x}$ and $R^*_{E_x^{(L)}}$ [Fig.~\ref{fig1}(d)]. The cancellation of positive $R^*_{E_x^{(L)}}$ by negative $R^*_{W_x}$ indicates that $E_x^{(L)}$ delays phase-locking, preventing premature acceleration at the wakefront where the ponderomotive force of $E_r^{(L)}$ would otherwise disperse the positrons.
Once $R^*<1$, the positron is phase-locked: $E_{lx}$ drives the longitudinal energy gain, while $R^*_{W_y}$ vanishes as the transverse velocity decreases to zero. Restoring the superluminal phase velocity ($v_{\rm ph}=1.05\,c$) introduces large-amplitude $R$ oscillations: the positron periodically slips across laser cycles, with $\bm v\times\bm B_\phi^{(L)}$ retrapping it into successive phases and imprinting an oscillatory modulation on $R$ [Fig.~\ref{fig1}(d)].}

\textcolor[rgb]{0.00,0.0,0.00}{The angular sensitivity of trapping is also captured by Eq.~(\ref{eq_Rpsi}). At fixed $p_0=4~m_ec$, decreasing $\theta_{\rm in}$ from $\pi$ toward $\pi/2$ shortens the $R^*<1$ phase window: more oblique incidence enhances the transverse momentum and thus $R^*_{W_y}$, ultimately ejecting the positron from the accelerating phase  [Fig.~\ref{fig1}(e)]. Head-on collisions therefore yield the most robust trapping. The momentum evolution of trapped positrons is provided in \cite{supplement}.}

\paragraph*{Generation of ultrabrilliant positron beams ---}{\color{black}Guided by the test-particle predictions, we perform full 3D PIC simulations using \emph{EPOCH} \cite{arber2015Contemporary,wan2023Simulations}. The simulation box ($25\lambda\times60\lambda\times60\lambda$, $800\times360\times360$ cells) moves along $\hat{\bm x}$ at $c$. Laser~1 is a tightly focused RPL described by the truncated-series fields ~\cite{salamin2006Accurate}.} \textcolor[rgb]{0.00,0.0,0.00}{To enhance superponderomotive injection against relativistic backward-directed positrons, its power is increased to $P=20$~TW (other parameters as in the test-particle model) and the plasma density is correspondingly raised to $n_0>0.01\,n_c$, with a flat-top profile between $x_1=30~\mu$m and $x_2=80~\mu$m and Gaussian ramps (with characteristic length 10 $\mu$m).} {\color{black}The pair jet is modeled by the \emph{EPOCH} injector module following estimated parameters \cite{chen2023Perspectives}: a J\"uttner spectrum of temperature 3~MeV, a divergence angle of $0.38$~rad, and a transverse Gaussian density profile of characteristic size $8~\mu$m with central density $10^{17}~\mathrm{cm^{-3}}$ at the converter rear surface.}

{\color{black}To verify the fundamental injection principle of the wake-jet collision via PIC simulation, the pair jet is set to head-on incidence ($\theta_{\rm in}=\pi$). Positrons entering the donut wake encounter a transverse field $W_y$ whose cyclic spatial structure, imprinted by $E_x^{(L)}$ on the merged inner sheath, acts as a spatial filter---only those crossing the near-axis focusing region are captured into accelerating phases of $E_{lx}$ [Fig.~\ref{fig2}(a)]. Particle tracking against the evolving $E_{lx}$ [Fig.~\ref{fig2}(b)] reveals that positrons starting at adjacent positions slip backward through successive laser cycles and lock into distinct accelerating phases $(\eta_{b1},\eta_{b2},\eta_{b3},\eta_{b4})$ once the longitudinal trapping occurs when $v_x/c\simeq1$. This phase-partitioning originates from the momenta spread of the pair jet: different $p_0$ and $\theta_{\rm in}$ produce distinct anti-dephasing trajectories and phase delays $\Delta\psi$, consistent with the $R^*(\psi)$ analysis [Figs.~\ref{fig1}(d)--(e)]. 
The averaged dephasing rate $\overline{R}$, normalized by initial one $\overline{R}_0$ of each phase-locked bunch converges toward $R\ll1$ nearly monotonically [Fig.~\ref{fig2}(c)]. The large-amplitude $R$-oscillations of individual positrons, driven by $v_\perp B_\phi^{(L)}\hat{x}$ as predicted by test-particle model [see Fig.~\ref{fig1}(d)], are largely averaged out within each bunch.}


{\color{black}The superponderomotive injection mechanism relies on the longitudinal wakefield strength, which can be tuned by a plasma density. At $n_0=0.025\,n_c$, $E_x^{(L)}$ dominates $E_{lx}$, imprinting a sharp laser-period modulation on the phase-space distribution that directly reflects the phase-partitioned trapping [Fig.~\ref{fig3}(a1)]. As $n_0$ increases to $0.05\,n_c$, $W_x$ overtakes $E_x^{(L)}$, smearing the discrete bunch structure while the peak current rises to $\sim1$~kA; a quasi-monoenergetic feature emerges at the current peak, indicating that the wakefield-dominated regime enforces more coherent acceleration [Fig.~\ref{fig3}(a3)]. The angular distributions exhibit tight collimation across all densities, with normalized emittance $\epsilon_{e^+}=(0.055,0.052,0.058)$~mm~mrad at $n_0=(0.025,0.035,0.050)\,n_c$ [Figs.~\ref{fig3}(b1)--(b3)]. This persistent low emittance arises from the periodic near-axis focusing by $W_y$ [Fig.~\ref{fig2}(a)], combined with the superponderomotive mechanism that channels energy gain into the longitudinal momentum, suppressing transverse divergence.} \textcolor[rgb]{0.00,0.0,0.00}{Note that the threefold symmetry observed in the angular distributions originates from the azimuthally non-uniform structure of $W_\perp$; a detailed analysis is provided in \cite{supplement}.}

{\color{black}The injected positron charge $Q_{e^+}$ is governed by geometric overlap, which is tuned by the incident angle $\theta_{\rm in}$ and the collision length $L_{\rm jet}$ [Figs.~\ref{fig3}(c)--(d)]. 
For fixed $L_{\rm jet}=40$ $\mu\mathrm{m}$, varying $\theta_{\rm in}$ changes both the incident positron momentum along $\hat{\bm{x}}$ and the wake--jet interaction distance.} \textcolor[rgb]{0.00,0.0,0.00}{Near perpendicular incidence ($\theta_{\rm in}\simeq\pi/2$), the interaction distance is minimized and the incident positrons carry larger transverse momentum, which reduces the trapping delay phase [see Fig.~1(e)], leading to minimal injected charge;} {\color{black}the angular spread of the trapped beam also varies with $\theta_{\rm in}$ through the phase-partitioning of the trapping process. Consequently, five-dimensional brightness $\mathcal{B}_{5\mathrm{D}}^+\propto Q_{e^+}/\epsilon_{e^+}$ is jointly governed by the charge and the emittance, both of which vary with $\theta_{\rm in}$. For fixed $\theta_{\rm in}=5\pi/6$, $Q_{e^+}$ is determined solely by $L_{\rm jet}$ and increases monotonically with it, while the emittance remains nearly constant. Hence $\mathcal{B}_{5\mathrm{D}}^+$ rises monotonically with the injected charge.} 
\textcolor[rgb]{0.00,0.0,0.00}{The broad density-angular tolerance and $L_{\mathrm{jet}}$-linear scalability together establish the robustness of superponderomotive injection.} 

{\color{black}The superponderomotive-injected beam, with its femtosecond duration and ultralow emittance, is well-suited for external injection into a hollow-channel plasma wakefield for GeV-level acceleration. This $\sim$0.05~mm~mrad emittance is three orders of magnitude smaller than the $\sim$50~mm~mrad attainable by damping-ring compression of BH positrons \cite{lindstrom2018Measurement,gessner2023Acceleration}, thereby substantially mitigating the transverse defocusing loss caused by beam misalignment in hollow-channel schemes. To verify that this beam also satisfies the more stringent size and divergence requirements of hollow-laser-driven donut-wake acceleration, we perform 3D PIC post-acceleration of the trailing beam obtained at $n_0=0.05\,n_c$ [Fig.~\ref{fig3}(a3)] in an LG-laser-driven donut wakefield. The driving LG laser satisfies the self-guiding condition $k_p w_0 \simeq 2\sqrt{a_0}$ with $a_0=9$ and $w_0 = 12\,\mu\mathrm{m}$, and a parabolic plasma channel ($n_e/n_c = 0.004 + 0.1\,r^2/w_0^4$) is employed. At early stage of post-acceleration, the beam is confined by the transverse wakefield $W_y$ while residing in the accelerating phase of $E_x$ [Figs.~\ref{fig4}(a)-(c)]. 
By $t=1300\,T_0$, $\sim$60\% of the positrons survive---edge positrons of the threefold structured trailing beam are defocused---while the retained beam exhibits substantial energy gain with reduced emittance ($\epsilon_{e^+}\simeq0.015$~mm~mrad); 
the leading two-cycle modulation arises as the beam front catches up with the laser field [Fig.~\ref{fig4}(d)]. Concurrently, the energy spread decreases monotonically to below $10\%$ as the beam gains GeV energy [Fig.~\ref{fig4}(e)]. The combination of lower energy spread and emittance yields a six-dimensional brightness $\mathcal{B}^+_{6\mathrm{D}}\simeq3.5\times10^{15}~\mathrm{A/m^{2}}/0.1\%$, comparable to the brightest electron beams from state-of-the-art LWFA and conventional linac bunches \cite{wang2016HighBrightness,dimitri2014Electron}. }

{\color{black}
This post-acceleration verification confirms that the inherent-injection beam's transverse and longitudinal characteristics can be well matched to the donut wakefield. Combined with established plasma channel-guiding technology \cite{gonsalves2019Petawatt,picksley2024Matched}, hollow-laser-driven wakefields such as LG lasers thus enable multi-GeV LWFA positron acceleration with percent-level energy spread. Experimentally, the requisite hollow laser pulses are within reach: LG pulses up to $10^{20}~\mathrm{W\,cm^{-2}}$ via reflection of a Gaussian beam through a high-reflectivity phase plate \cite{wang2020Hollow,zhang2022Ultraintense,wang2026Experimental}, and terawatt RPL pulses via high-gain parametric amplification \cite{kong2019Generating,powell2024Relativistic,zhong2021Polarizationinsensitive}.}

{\color{black}In summary, we have proposed a superponderomotive injection method for inherent positron trapping in the blowout regime of LWFA. The mechanism exploits laser--wakefield synergy to suppress the dephasing rate, as captured by a phase-integral equation. 3D PIC simulations demonstrate this scheme via donut-wake--pair-jet collision, generating ultralow-emittance multicycle positron beams. Start-to-end simulations confirm high-throughput injection-to-acceleration coupling, with post-acceleration yielding quasi-monoenergetic beams of six-dimensional brightness $\sim10^{15}~\mathrm{A/m^{2}}/0.1\%$, thus resolving the key injection-beam-quality challenge in plasma-based positron acceleration.  
Building on currently feasible hollow laser technology and mature BH pair-jet sources, this scheme opens a novel compact route to ultrabrilliant positron beams, with immediate relevance to ultrafast material diagnostics, laboratory modeling of pair-driven astrophysical environments, and the development of next-generation electron--positron colliders.}


\vskip 0.5cm
{\it Acknowledgements---{\rm The}} work is supported by the National Natural Science Foundation of China (Grants No. 12425510, No. U2267204, No. 12441506, No. 12475249, No. 12447106, No. 12275209), the Science Challenge Project (No. TZ2025012), the National Key Research and Development (R\&D) Program (Grant No. 2024YFA1610900, No. 2024YFA1612700), the Innovative Scientific Program of CNNC, Natural Science Basic Research Program of Shaanxi (Grant No. 2024JC-YBQN-0042), and the Fundamental Research Funds for Central Universities (No. xzy012023046).

\bibliography{refs_positron}

\end{document}